\documentclass{aa}
\usepackage{epsfig,amsfonts,amssymb}
\begin{document}

\thesaurus{10.08.1, 10.11.1, 10.19.2, 12.04.1, 12.07.1}

\title{Microlensing towards the Small Magellanic Cloud\\
EROS 2 two-year analysis
\thanks{Based on observations made at the European Southern Observatory,
La Silla, Chile.}}
\author{
C.~Afonso\inst{1}, 
C.~Alard\inst{9},
J.N.~Albert\inst{2},
J.~Andersen\inst{6},
R.~Ansari\inst{2}, 
\'E.~Aubourg\inst{1}, 
P.~Bareyre\inst{1,4}, 
F.~Bauer\inst{1,4},
J.P.~Beaulieu\inst{5},
A.~Bouquet\inst{4},
S.~Char\inst{7},
X.~Charlot\inst{1},
F.~Couchot\inst{2}, 
C.~Coutures\inst{1}, 
F.~Derue\inst{2}, 
R.~Ferlet\inst{5},
J.F.~Glicenstein\inst{1},
B.~Goldman\inst{1},
A.~Gould\inst{1,8}\thanks{Alfred P.\ Sloan Foundation Fellow},
D.~Graff\inst{8},
M.~Gros\inst{1}, 
J.~Haissinski\inst{2}, 
J.C.~Hamilton\inst{4},
D.~Hardin\inst{1},
J.~de Kat\inst{1}, 
A.~Kim\inst{4},
T.~Lasserre\inst{1},
\'E.~Lesquoy\inst{1},
C.~Loup\inst{5},
C.~Magneville \inst{1}, 
B.~Mansoux\inst{2}, 
J.B.~Marquette\inst{5},
\'E.~Maurice\inst{3}, 
A.~Milsztajn \inst{1},  
M.~Moniez\inst{2},
N.~Palanque-Delabrouille\inst{1}, 
O.~Perdereau\inst{2},
L.~Pr\'evot\inst{3}, 
N.~Regnault\inst{2},
J.~Rich\inst{1}, 
M.~Spiro\inst{1},
A.~Vidal-Madjar\inst{5},
L.~Vigroux\inst{1},
S.~Zylberajch\inst{1}
\\   \indent   \indent
The EROS collaboration
}
%% 1 Saclay, 2 LAL, 3 Marseille, 4 CdF, 5 IAP, 6 Copenhague, 7 La Serena,
%% 8 Gould & Graff, 9 Alard, 
\institute{
CEA, DSM, DAPNIA,
Centre d'\'Etudes de Saclay, 91191 Gif-sur-Yvette Cedex, France
\and
Laboratoire de l'Acc\'{e}l\'{e}rateur Lin\'{e}aire,
IN2P3 CNRS, Universit\'e Paris-Sud, 91405 Orsay Cedex, France
\and
Observatoire de Marseille,
2 pl. Le Verrier, 13248 Marseille Cedex 04, France
\and
Coll\`ege de France, Physique Corpusculaire et Cosmologie, IN2P3 CNRS, 
11 pl. M. Berthelot, 75231 Paris Cedex, France
\and
Institut d'Astrophysique de Paris, INSU CNRS,
98~bis Boulevard Arago, 75014 Paris, France
\and
Astronomical Observatory, Copenhagen University, Juliane Maries Vej 30, 
2100 Copenhagen, Denmark
\and
Universidad de la Serena, Facultad de Ciencias, Departamento de Fisica,
Casilla 554, La Serena, Chile
\and
Departments of Astronomy and Physics, Ohio State University, Columbus, 
OH 43210, U.S.A.
\and
DASGAL, 77 avenue de l'Observatoire, 75014 Paris, France
}
\offprints{Nathalie.Delabrouille@cea.fr}

\date{Received;accepted}

\authorrunning{C. Afonso et al.}
\titlerunning{EROS~2 SMC two-year analysis}

\def\kms{{\rm km}\,{\rm s}^{-1}}
\def\kpc{{\rm kpc}}
\def\lsim{{\lesssim}}
\def\au{{\rm AU}}
\def\etal{{et al.}}
\def\eros{{\sc eros}}
\def\macho{{\sc macho}}
\def\lmc{{\sc lmc}}
\def\smc{{\sc smc}}
\def\ie{{\em i.e.}}

\maketitle

\begin{abstract}

We present the analysis of the first two years of 
a search for microlensing of stars in
the Small Magellanic Cloud with the \eros\ (Exp\'erience de Recherche
d'Objets Sombres) project. A single event is detected, already present
in the first year analysis. This low event rate allows us to put new
constraints on the fraction of the Galactic Halo mass due to compact
objects in the mass range [$10^{-2}$,
$1$]$\:{\rm M}_{\odot}$. These limits, along with the fact that the two 
\smc\ events observed so
far are probably due to \smc\ deflectors, suggest that  \lmc\
and \smc\ self-lensing may dominate the event rate.


\end{abstract}

\section{Introduction}
	
	In 1986, B. Paczy\'nski (\cite{Pac86}) proposed to use
gravitational microlensing as a tool to probe the dark matter of the
Galactic Halo. A few years later, several experiments started looking
for microlensing events towards the Large Magellanic Cloud
(\lmc). Several candidates have now been observed by the \macho\ and
\eros\ collaborations 
(\cite{alc93, aub93}, Alcock et al. 1997a, \cite{ans96}).

The first results towards the \lmc\ suggested an optical depth of 
order half that required to account for the dynamical mass of the dark 
halo, and the typical Einstein radius crossing time of the events 
implied a mass of about 0.5~M$_\odot$ for the lenses (\cite{alc97a}).  
The results of the \eros\ photographic plates microlensing survey
pointed to a somewhat lower optical depth (Ansari et al. 1996).
Such objects cannot be main-sequence stars, whose observed density is 
two orders of magnitude too low to account for the measured optical 
depth.  Many hypotheses have been debated regarding the nature and 
location of these lenses, as LMC self-lensing (\cite{sahu94, wu94}), 
tidal debris (\cite{zhao98}), intervening populations (Zaritsky \& Lin 
1997), a warped Galactic Disk (Evans et al.  1998), white dwarfs 
(Adams \& Laughlin 1996, Graff et al. 1998), primordial black holes 
(Jedamzik 1998), all 
of which have brought forth numerous counter-arguments (Gould 1995, 
Alcock et al.  1997b, Gould 1998, Bennett 1998, Beaulieu \& Sackett 
1998).
No fully satisfactory possibility has emerged, the
main two trends being 0.5~M$_\odot$ dark objects in the halo, 
such as primordial black holes, 
or 0.1~M$_\odot$ main sequence stars near the \lmc\ itself. The
cosmological implications of these two hypotheses differ considerably.

	The Small Magellanic Cloud (\smc) is highly valuable as a second
target for halo microlensing searches. If the lenses belong to
the halo of our Galaxy, the optical depth and the distribution of
event durations should be similar to the ones observed towards the
\lmc. In the case of \smc\ or \lmc\ self-lensing, however, the
different dynamical properties of the two Clouds will give rise to
differences in optical depth and event duration.

	{\sc Eros}, whose setup was upgraded in 1996, is engaged
in observations towards the \lmc\ and the \smc.  The analysis of the
first year of observations towards the \smc\ yielded one event
(Palanque-Delabrouille et al. 1998, Alcock et al 1997c).  A second event,
whose source star was too faint to be in the \eros\ reference catalog,
was alerted on by \macho\ and actively monitored by all microlensing
collaborations. It was generated by a binary deflector most probably
located in the \smc\ itself (\cite{af98}, Albrow et al. 1999, Alcock
et al. 1999, Rhie et al. 1999, Udalski et al. 1999). This light curve, 
however, is not counted
as an event in the following analysis which only uses stars identified
on our template images, for which the efficiency can be estimated.

We present here the details of the \eros\ 2 two-year analysis of data
towards the \smc\, and discuss its implications for the nature of the
halo dark matter.

\section{Experimental setup and observations}

	The telescope, camera, telescope operations and data reduction
are as described in Palanque-Delabrouille \etal\ 1998 and references
therein. Ten square degrees are monitored on the \smc. Note that since
one {\sc ccd} of the red camera was non functional, all the analysis
is done on only 7 {\sc ccd}'s (\ie\ a total field of 8.6 square degrees).

The two-year data set contains 5.3 million light curves covering the
period from July 1996 to March 1998. 

\section{Data analysis}

The analysis of the two-year data set is similar to that of the first
year (Palanque-Delabrouille et al. 1998). The major difference comes
from the addition of rejection cuts requiring a minimum time coverage of
the event (not possible with the previous one-year time span), and
from the tuning of some cuts applied to parameters with a clear
field-dependent distribution. The criteria are
sufficiently loose not to reject light curves deformed by blending or by the
finite size of the source, or events involving multiple lenses or
sources. Most variable stars are rejected by at least two distinct
cuts.

As in the first year analysis, we define a positive (negative)
fluctuation as a series of data points that (i) starts by one point
deviating by at least $1\sigma$ from the baseline flux $\phi_{0}$, (ii) stops with at
least three consecutive points below $\phi_{0} + 1\sigma$
(above $\phi_{0} - 1\sigma$) and (iii) contains at least 4
points above $\phi_{0} + 1\sigma$ (below $\phi_{0} -
1\sigma$). The significance $LP$ of a given variation is defined as:
\begin{equation}
 LP = - \sum_{i=1}^{i=N}
   \log\left(\frac{1}{2}\:{\rm Erfc}\left(\frac{x_{i}}{\sqrt{2}}\right)\right) 
\end{equation}
where $x_i$ is the deviation in $\sigma$'s of the point taken at time
$t_i$ and $N$ the number of points within the fluctuation.  We order
the fluctuations along a light curve by decreasing significance. The
cuts are described hereafter:
\vspace{.5cm}\\ 
\noindent{\bf Selection of microlensing candidates:}
\begin{itemize}
\item {\bf 1a:} The main fluctuation detected in the red and blue
light curves should be positive and occur simultaneously in the two
colors: if I is the time interval during which the data are more than
$1\sigma$ away from the baseline flux, we require $({\rm I_{red}} \cap
{\rm I_{blue}})/({\rm I_{red}} \cup {\rm I_{blue}})>0.2$.
\item {\bf 1b:} To reject flat light curves with only statistical
fluctuations, we require that on a given light curve \\ $LP({\rm
2^{nd}\:most\:significant\:fluct.})\:/\:LP({\rm main\:fluct.})<0.6$\\
in both colors.\footnote{This criterion has been relaxed slightly
compared to the first year analysis.}
\item {\bf 1c:} We require that $LP(\rm main\;fluct.)>30$ in both colors.
\end{itemize}
{\bf Rejection of variable stars:}
\begin{itemize}
\item {\bf 2a:} To exclude short period variable stars that exhibit
irregular light curves, we require that the RMS of the distribution of
the deviation, in $\sigma$'s, of each flux measurement from the linear
interpolation between its two neighboring data points be small,
typically less than 2.2 (the exact value varying from field to field,
with a looser limit set on outer fields).
\item {\bf 2b:} To exclude variable stars which exhibit correlated
fluctuations between the two colors, we calculate the correlation
coefficient $\rho$ between the ``red'' and the ``blue'' light curves,
excluding points in the main fluctuation (enlarged by a 25\% time
margin on either side) so as to consider only the un-amplified part of
the light curve. The cut is applied to 
%\[{\rm cor} = \frac{\rho}{\sqrt{1-\rho^2}}*\sqrt{N_\rho-1}\]
\[{\rm cor} = \sqrt{N_\rho-2} \times {\rho}/{\sqrt{1-\rho^2}}\] whose
distribution is less sensitive than $\rho$ to the number of points
$N_\rho$.  We require ${\rm cor} < 6$.
\item {\bf 2c:} We remove two under-populated regions of the
color-magnitude diagram that contain a large fraction of variable
stars (upper main sequence and bright red giants, see details in 
first year analysis).
\end{itemize}
{\bf S/N improvement of the set of selected candidates:}
\begin{itemize}
\item {\bf 3a:} We remove events with low signal-to-noise ratio by requiring
a significant improvement of a microlensing fit (ml) over a constant
flux fit (cst), i.e. that \\ $[\chi^2({\rm cst}) -\chi^2({\rm
ml})]/[\chi^2({\rm ml})/{\rm d.o.f.}]> \Delta \chi^2_{\rm min}$ \\
where d.o.f. is the number of degrees of freedom. As for cut 2a, the
exact value of $\Delta \chi^2_{\rm min}$ varies from field to field
because of the variations in time sampling. Typically, $\Delta \chi^2_{\rm
min} \sim 350$, with a higher limit on well-sampled
fields.
\item {\bf 3b:} We require that the maximum magnification in the
microlensing fit be greater than 1.40.
\end{itemize} 
{\bf Time coverage of the event:} 
\begin{itemize} 
\item {\bf 4a:} We require that the fitted time of maximum magnification be
contained in the period of observation.
\item {\bf 4b:} We require that the fitted value of the Einstein
radius crossing time $\Delta t<300$ days.
\end{itemize}
{\bf Physical blending:}
\begin{itemize}
\item {\bf 5:} If blending significantly improves the microlensing
fit, then the fitted blending should be physical, \ie: if $\chi^2({\rm
no\;bl.})-\chi^2({\rm bl.})>10$, we require that the fitted
blending fraction $f_{\rm bl} / f_{\rm tot} > -0.3$, in both colors.
\end{itemize}
The tuning of each cut and the estimation of the efficiency of the
analysis (with the correction due to blending) is done with Monte
Carlo simulated light curves, as described in Palanque-Delabrouille 
et al. 1998. The
microlensing parameters are drawn uniformly in the following
intervals: time of maximum magnification $t_0 \in [t_{\rm first}-350,
t_{\rm last}+350]$ days, impact parameter normalized to the Einstein
radius $u_0 \in [0,2]$ and time-scale 
$\Delta t \in [0,350]$ days.  Cuts 2a and 3a are tuned individually on
each field so as to reject respectively $\sim$~15\% and $\sim$~25\% of
the remaining Monte Carlo light curves. The impact of each cut on data
and simulated events is summarized in table~1.
\vspace{-.3 cm}
%\vspace{-.5 cm}
\begin{table}[ht]
\begin{center} \vspace{-0.1cm} \begin{tabular}{|l||c|c|c|} 
\hline
 &{\em Number of}& \multicolumn{2}{c|}{\em Fraction of remaining} \\
{\em Cut description}&{\em stars}&\multicolumn{2}{c|}{\em stars removed by cut}
\\ \cline{3-4}
& {\em remaining} & {\em \ \ \ Data\ \ \ } & {\em Simulation}\\
\hline
Stars analyzed & 5,307,774 & - & - \\
1a: Simultaneity & 93,900&  98\% &76\%\\
1b: Uniqueness  &  29,742&  68\% &11\%\\
1c: Significance &  9,978&  66\% &06\%\\
2a: Stability &     4,365&  56\% &16\%\\
2b: Correlation &   3,502&  20\% &02\%\\
2c: HR diagram &    2,996&  14\% &04\%\\
3a: Microlensing fit &102&  97\% &24\%\\
3b: Magnification &    37&  63\% &16\%\\
4a: Time inclusion &   28&  24\% &20\% \\
4b: Duration &         22&  21\% &14\% \\
5: Blending &           1&  95\% &12\%\\
\hline
\end{tabular}
\caption{Impact of each cut on data and simulated microlensing events
(allowing for blending). Each fraction for cut $n$ refers to the light
curves remaining after cut $(n-1)$.}\label{cuts}
\vspace{-0.5cm} 
\end{center}
\end{table}
\vspace{-.5cm} 

\begin{figure} [h] 
 \begin{center} \epsfig{file=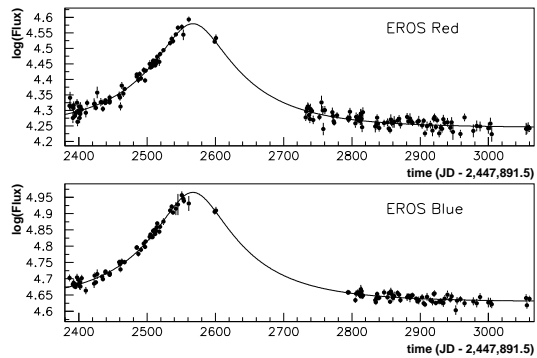,width=7.8cm} \vspace{-.1cm}
  \caption{Light curve of microlensing event EROS-SMC-97/1 (the fit shown
  does not include the periodic modulation).}
% Time is in days since
%   Jan. 0, 1990 (JD 2,447,891.5).}  
  \label{cdl_evt}
 \end{center} %\vspace{-0.7cm}
\end{figure}

There is one remaining candidate, the one already discussed in the
analysis of the first year data set. Its updated light curve is shown
in figure~\ref{cdl_evt}.

The efficiency of the analysis for events with an impact parameter
$u_0<1$ and normalized to an observing period $T_{\rm obs}$ of one
year is summarized in table~\ref{eff}, as a function of the Einstein
radius crossing times $\Delta t$ (in days). The main source of
systematics could come from a biased estimation of the blending
effect. We have studied various blending models and found the relative
error to be less than 10\%. 
%Figure~\ref{nbevt} illustrates the
%increase in the number of expected events as compared to the first
%year analysis. This new analysis being more strict than before (in
%particular cuts 4a and 4b), the number of events expected is less than
%twice that of the previous year.
\begin{table}[ht]
%\vspace{-0.3cm}
 \begin{tabular}{|c||c|c|c|c|c|c|c|c|c|c|} \hline
$\Delta t$ & 7 & 22&  37 & 52& 67& 100& 150& 250& 300& 350\\ \hline
$\epsilon$ & 4 & 12&  16 & 18& 21&  24&  25&  23&  13&   3\\ \hline
\end{tabular}
\caption{Efficiency (in \%) of the analysis as a function of the
time-scale $\Delta t$ (in days) for events with $u_0<1$, normalized
to $T_{\rm obs}=1{\rm \: yr}$. We monitor $N_{\rm obs}=5.3\times 10^6$
stars.}
\label{eff} \vspace{-0.6cm}
\end{table}

\section{Limits on the contribution of dark compact objects to the Halo} 
We fit a microlensing curve to the data of the candidate (see table
\ref{eventparm}), allowing for a periodic modulation of the source
star as evidenced in the first year analysis. Blending and time-scale
being degenerate quantities whose common fit is quite sensitive to
systematics, we set the blending to that measured by {\sc ogle}
(\cite{ud97}): $f_{\rm bl} / f_{\rm tot} \sim 24\%$, considering it to
be a lower limit on the actual contribution (due to possible
additional non-resolved companions). This value of the blending
fraction is compatible with the absence of distorsion in the spectrum
(\cite{sahu98}).

\begin{table}[ht]
\begin{center} \vspace{-0.1cm} \begin{tabular}{|c|c|c|c|c|c|}
\hline
$u_0$      &  $t_0$    & $\Delta t$& $Period$ & $A_{\rm mod\;R}$ & $A_{\rm mod\;B}$ \\
\hline
$0.424$     & $2568.8$  & $129$      & $5.126$     & $0.031$    & $0.022$ \\
$\pm 0.004$ & $\pm 0.8$ & $\pm 2$  & $\pm 0.002$ & $\pm 0.003$& $\pm 0.003$\\
\hline
\end{tabular}
\caption{Result of microlensing fit to the \smc\ event, with blending
set to $f_{\rm bl}/ f_{\rm tot}= 25\%$. $t_0$ is the time of maximum
magnification in days since Jan, 1 1990, $\Delta t$ the Einstein
radius crossing time, in days. The $\chi^2$ is 261 for 279
d.o.f.}
\label{eventparm}
\end{center} \vspace{-0.5cm}
\end{table}

In order to set limits on the contribution of dark objects to the
Halo, a Halo model must be used to obtain, for a given deflector mass,
both a number of expected events and a distribution of event
durations.  We use the so-called ``standard'' halo model described in
Palanque-Delabrouille et al. 1998 as model 1, and take into account
the efficiency of the analysis given in the previous section.

Moreover, to set conservative limits, we have assumed
that the observed event is a halo event, in spite of the fact that it
is most probably an \smc\ self-lensing event (Palanque-Delabrouille et
al. 1998).

Assuming a standard halo model with a mass fraction $f$ composed of
dark compact objects having a single mass $M$, the likelihood of
observing at most one event, with a duration less probable than the
observed one ($\Delta t_{\rm obs}$), is\vspace {-.1cm}
\begin{equation}
L(M) = e^{-f \tilde{N}_M} \left( 1 + f \tilde{N}_M 
       {\cal P}_M(\Delta t_{\rm obs})  \right)
\end{equation}%\vspace {-.2cm}
with %\vspace {-.2cm}
\begin{equation}
{\cal P}_M(\Delta t_{\rm obs}) = \int_0^{\Delta t_1} P_M(\Delta t) d\Delta t + 
                             \int_{\Delta t_2}^\infty P_M(\Delta t) d\Delta t 
\end{equation}
$\Delta t_1$ and $\Delta t_2$ are the two durations with the same 
probability as $\Delta t_{\rm obs}$, defined by:
%\vspace {-.2cm}
\begin{eqnarray}
P_M(\Delta t_1) = P_M(\Delta t_2),\: \Delta t_1<\Delta t_2,\: 
\nonumber \\ 
(\Delta t_1 = \Delta t_{\rm obs} \; \mathrm{or}\; \Delta t_2 = \Delta t_{\rm obs})
\end{eqnarray}%\vspace {-.1cm}
$P_M(\Delta t)$ is the distribution of expected event durations, taking efficiencies
into account, and $\tilde{N}_M$ is the expected number of 
events.\footnote{In Palanque-Delabrouille et al. 1998, we derived a
constraint on parallax. Using either only the constraint on the
observed duration as we do here, only the constraint on parallax, or
both constraints simultaneously, yield equivalent exclusion limits.}

\begin{figure} [h] 
 \begin{center} \epsfig{file=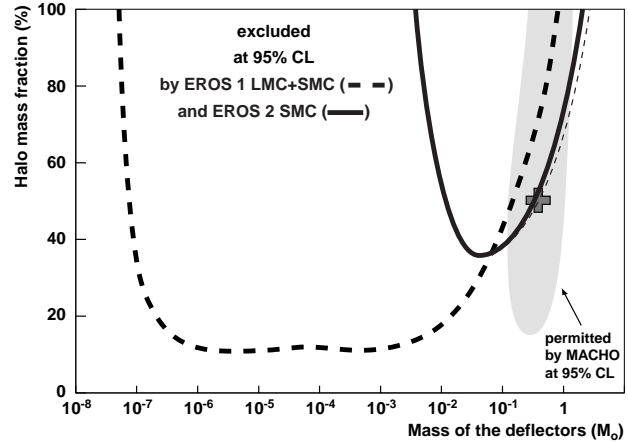,width=8.7cm}
  \vspace{-.1cm} \caption{ Exclusion diagram at 95~\% C.L. for the
  standard halo model ($4 \times 10^{11}\:{\rm M}_\odot$ inside
  50~kpc).  The dashed line is the limit from \eros\ 1 \lmc\ and \smc\
  data (Renault et al., 1997), the solid line is the limit from the
  \smc\ data described in this paper.  The {\sc macho} 95~\%
  C.L. accepted region is the shaded area, with the preferred value
  indicated by the cross (Alcock et al. 1997a). The thin dashed line
  corresponds to the limit obtained assuming we observed no halo 
  events. } \label{excl}
  \end{center} \vspace{-1cm}
\end{figure}

Figure \ref{excl} shows the 95~\% exclusion limit derived from this 
likelihood function on $f$, the halo mass fraction,
at any given mass $M$, \ie\ assuming all deflectors in the halo have mass $M$.
For comparison, the exclusion curve obtained considering that
the sole event detected does not belong to the halo (thin dashed line 
on the plot) is also shown. The limit is very similar to the one event 
one, since the very long duration of the event pushes it towards large
masses, where our efficiency starts to drop.

\section{Discussion and conclusion}
The analysis of two years of \eros\ 2 \smc\ data has yielded a single
microlensing event. This allows us to put new constraints on the fraction
of the halo made of objects in the range [$10^{-2}\:{\rm M}_\odot$,
$1\:{\rm M}_{\odot}$], excluding in particular at the 95~\% C.L. that more
than 50~\% of the standard halo be made of $0.5 \:{\rm M}_\odot$ objects.

One should also note that all the microlensing events towards the
\lmc\ and the \smc\ for which information could be obtained on the
deflector location --- through parallax for this 97-SMC-1 event
(Palanque-Delabrouille et al. 1998), and deflector binarity for
98-SMC-1 and LMC-9 (but see the discussion in Bennett et al. 1996) ---
seem to be produced by deflectors located in the Clouds
themselves. This suggests that self-lensing may be the dominant source
of the observed events. We are continuing to accumulate data towards
the \lmc\ and \smc\ in order to determine definitively the location
of the deflectors.

\begin{acknowledgements}
We are grateful to D. Lacroix and the technical staff at the Observatoire de
Haute Provence and to A. Baranne for their help in refurbishing the MARLY
telescope and remounting it in La Silla. We are also grateful for the support
given to our project by the technical staff at ESO, La Silla. We thank
J.F. Lecointe for assistance with the online computing, and
J. Bouchez for useful comments.
\end{acknowledgements}

\end{document}